\documentclass[preprint,12pt]{elsarticle}
\usepackage{amssymb}
\usepackage{amsmath}
\usepackage{xcolor}
\usepackage{hyperref}
\usepackage{caption}
\usepackage{subcaption}
\usepackage{soul}

\newcommand{\uu }{\boldsymbol{u}}
\newcommand{\nab }{\boldsymbol{\nabla}}

\newcommand{\Ra}{\text{Ra}}
\newcommand{\Nus}{\text{Nus}}

\journal{Communications in Nonlinear Science and Numerical Simulation}

\begin{document}

\begin{frontmatter}



\title{Dynamics and Scaling of Internally Cooled Convection}


\author[aff1]{Lokahith Agasthya}
\author[aff1]{Caroline Jane Muller}

\affiliation[aff1]{organization={Institute of Science and Technology Austria},%Department and Organization
            addressline={Am Campus 1}, 
            city={Klosterneuburg},
            postcode={3400}, 
            country={Austria}}

\begin{abstract}
%\st{We study  the system of internally cooled convection first introduced by Berlengiero et. al.}
%\ca{{\it this first sentence is a little vague (see also comment at the beginning of the introduction). Maybe this can be made more precise on goals and key points/results, e.g.} "Our goal is to investigate  fundamental properties of the system of internally cooled convection. The system resembles classical Rayleigh Benard convection, but with an interior energy sink represented by a constant cooling term, interpreted as atmospheric radiative cooling. This work builds upon and expands recent work by Berlengiero et. al. on internally cooled convection, and addresses fundamental questions, such as the derivation of a scaling for the convective mass flux, and the breakage of up/down symmetry for convection in the presence of internal cooling." }

Our goal is to investigate fundamental properties of the system of internally cooled convection. The system consists of an upward thermal flux at the lower boundary, a mean temperature lapse-rate and a constant cooling term in the bulk with the bulk cooling in thermal equilibrium with the input heat flux. This simple model represents idealised dry convection in the atmospheric boundary layer, where the cooling mimics the radiative cooling to space notably through longwave radiation. 
We perform linear stability analysis of the model for different values of the mean stratification to derive the critical forcing above which the fluid is convectively unstable to small perturbations. The dynamic behaviour of the fluid system is described and the scaling of various important measured quantities such as the total vertical convective heat flux and the upward mass flux is measured. We introduce a lapse-rate dependent dimensionless Rayleigh-number $\Ra_\gamma$ that determines the behaviour of the system, finding that the convective heat-flux and mass-flux scale scale approximately as $\Ra_\gamma^{0.5}$ and $\Ra_\gamma^{0.7}$ respectively. The area-fraction of the domain that is occupied by upward and downward moving fluid and the skewness of the vertical velocity are studied to understand the asymmetry inherent in the system. 
We conclude with a short discussion on the relevance to atmospheric convection and the scope for further investigations of atmospheric convection using similar simplified approaches. 
\end{abstract}


% \begin{highlights}
% \end{highlights}

%\begin{keyword}
%Keyword1, keyword2, keyword3
%\end{keyword}

\end{frontmatter}


\section{Introduction}
%\ca{My only major comment is to mention clearly the goals and key results of the study, and place them in the context of existing results/literature. e.g. study Rayleigh-Benard convection with interior cooling, which is relevant for atmospheric boundary layer convection (below cloud base where water phase changes, which are not included here, become important). Here we are interested in clarifying some fundamental aspects of this system. More precisely, the questions addressed here are - can we derive scalings for quantities of interest for atmospheric convection, including mass flux. How do these depend on model parameters?; - up/down symmetry breaking? upward convective area? Upward velocities? }

The study of thermal, convective systems has a long history going back to the late 19th century, with the recorded observations of James Thomson \cite{thomson1882changing} and the systematic experiments performed by Henri B\'enard \cite{benard1901} setting the stage for further investigation into the effects of heating a fluid. Lord Rayleigh was the first to analytically describe the convective instability resulting from heating a fluid from below \cite{rayleigh1916} using a model system that was named Rayleigh-B\'enard (R-B) convection and one that has become the bedrock of studies on the behaviour of thermal fluids and phenomena such as pattern formation, transition to chaos etc. \cite{ahlers2009heat,getling1998rayleigh}

One of the primary motivations to study R-B convection and other similar model fluid systems is to understand and characterise the behaviour of the Earth's atmosphere. The atmospheric circulation of the earth is driven primarily by the heating of the earth's surface by the sun, which in turn heats the lowest level of the atmosphere. In addition to simple thermal effects, atmospheric convection includes a large number of physical, chemical and biological processes at different scales \cite{emanuel1994atmospheric}.
The study of atmospheric convection is usually performed by solving a large set of coupled non-linear equations which represent all these processes. These could be General Circulation Models (GCM) at the global scale \cite{manabe1965simulated,edwards2011history} or Cloud Resolving Models (CRM) at smaller scales \cite{guichard2017short}. While they show realistic behaviour, their complexity, especially the large number of state variables in the models makes their results hard to interpret and can even obscure a more fundamental understanding of atmospheric processes. 

In the context of atmospheric convection, moist convection  (ie., convection in the presence of moisture) is one of the most important phenomenon in the tropics. Here, the release of latent heat due to phase changes in water is very important to the dynamics, while the micro-physical details of the nature, number and size of hydrometeors (water in liquid or solid form) have an important feedback on the convective scale dynamics. 
Recently several studies have focused on developing simplified models which ignore several processes at various scales but capture essential features of the real atmosphere - for some examples, see \cite{hernandez2013minimal,weidauer2012moist,vallis2019simple}. These studies considered moist convection with highly simplified cloud microphysics, moist convection with varying thermal boundary conditions and moist-convection without considering the dynamics of liquid water respectively. They found robust evidence of an up-down asymmetry (where hot, rising updrafts occupy a smaller fraction of the domain than cold, subsiding air) %downdrafts) %CM: downdrafts usually refer to fast downward air below precipitating clouds
and scalings with different parameters for important measured quantities such as the heat-flux, the mass flux in clouds, the height of maximum cloud formation, etc. Further exploration of such simplified convective models with some basic processes of the atmosphere represented while omitting or greatly simplifying other processes has the potential to uncover which are the fundamental processes which set the dynamics of tropical moist convection and which do not greatly affect the behaviour of moist convection. 

%\ca{\it Say a few words about each paper you cite (main assumptions / equations and key points. Like you do below is ok, maybe move the references there in the discussion?}
In this study, we focus on the simpler case of dry convection which occurs in the atmosphere when moisture is either absent or present in small enough quantities that condensation or freezing can be neglected. In other words, we consider convection of dry air or of moist air (i.e. with water vapor), but without phase change of water vapor into
liquid or ice. For simplicity we will derive the equations for dry air, but we note that water vapor could easily be accounted for by replacing temperature with virtual temperature \cite{emanuel1994atmospheric}.
Dry convection occurs in the region between the earth's surface and the cloud-base, known as the sub-cloud layer. 
Dry convection is prevalent in the tropics over dry land and is particularly important in studies of the planetary boundary layer. Here, the constant cooling of the atmosphere by the emission of longwave radiation plays a non-negligible role in the dynamics. 

To study this convective boundary layer, an idealisation that is most often made is to study the model fluid system with a fixed temperature at the lower boundary and (unlike Rayleigh-B\'enard convection) a constant rate of diabatic cooling everywhere in the domain. An account of the global-heat balance and scaling of the heat-flux for this system can be found in Chapter 3.6 of \cite{emanuel1994atmospheric}, while the results from more recent simulations are reported in \cite{hartlep2018convection}.

%An analysis of the stability of the system can be found in 

Berlengiero et. al. in their study \cite{berlengiero2012internally} (henceforth referred to as B2012) investigated the convective behaviour of a layer of fluid with a fixed heat-flux at the lower surface and a uniform bulk-cooling term to understand the basic features of atmospheric dry convection with a constant radiative cooling to space. They argued that in some scenarios, particularly over land, fixed-flux boundary conditions is a better representation of the heating of the atmosphere by the surface than fixed temperature.
The system consists of a layer of a Boussinesq fluid with a fixed input heat-flux $f_0$ at the lower surface, no vertical heat-flux at the upper boundary, a constant bulk-cooling $-R$ and a set adiabatic temperature lapse rate $\gamma$. The heat input into the system and the cooling via the bulk-cooling term must be in global balance to ensure that a system at thermal equilibrium is achieved. 

In B2012, the authors described the dynamics of two different 3D flows, one with a non-zero, finite temperature stratification $\gamma$ and a second flow with no stratification. The results presented included the vertical temperature profile for both cases, the volume fraction of the domain occupied by updrafts %updraughts %CM: Use same spelling as elsewhere  
and the skewness of the vertical velocity. In particular, it was found that for both cases, less than half the volume was occupied by upward velocities, with concentrated intense hot plumes and large regions of subsiding flow. It was also found that the flow with $\gamma = 0$ showed strong clustering of hot plumes due to the interaction between the rising plumes and the returning subsiding flow from the upper boundary. 

In the absence of a mean stratification ($\gamma = 0$ case), this set-up is equivalent and dynamically identical to other well-studied models of convection. Here we highlight the works of Goluskin \cite{goluskin2015internally,goluskin2016internally} and the work conducted jointly by Aumaître, Gallet and others \cite{lepot2018radiative,bouillaut2019transition,miquel2020role}. 
Goluskin considered a system of internally heated convection balanced by a constant outward heat-flux at the top boundary with an insulating lower boundary. This system is the same as the B2012 system after the transformations $z \to -z$ and $T \to -T$. 
Goluskin carried out the linear stability analysis of the system and inferred the critical non-dimensional forcing (Rayleigh number) for which the system is unstable to small perturbations. The works of Aumaitre, Gallet et. al consider a column of fluid that is cooled internally and heated internally. While the internal cooling extends throughout the domain, the heating extends from the lower surface up to a given
typical height $h$. When $h \to 0$, this is identical to a fixed thermal-flux boundary condition. They characterised the convective transport, the flow and plume patterns and the nature of the boundary layer for varying $h$, finding that depending on whether the heating is localised in the boundaries ($h \to 0$) or more spread across the domain ($h$ larger than typical height of the boundary layer), different regimes of turbulent transport are achieved.

% set-up has also been studied in \cite{lepot2018radiative}, albeit for a adiabatic lapse-rate $\gamma$ set to $0$. The focus in their work was on the difference in the nature of the boundary layer and resulting net heat transfer when the heating was distributed over a small height near the lower surface and not only the lower boundary. 

%\ca{\it Maybe discuss the boundary forcing of Rayleigh Benard versus atmospheric interior cooling, which thus can lead to different behavior. Also the bottom and top forcing of Berlengiero et al is in the form of surface flux (I think?) unlike fixed T in Rayleigh Benard. This is meant to mimic surface sensible heat flux at the bottom of the atmosphere near the surface.}

In the presence of stratification ($\gamma$ non-zero), the situation is more complicated given that the fluid instability extends only up to a finite height in the domain \cite{berlengiero2012internally}, with a layer of stable fluid over a layer of unstable fluid. This is the so-called ``capping inversion" that occurs in the atmosphere on fair-weather days where the well-mixed atmospheric boundary layer is capped by a layer of statically stable air \cite{capping1973}. In the ice-water system \cite{roberts1985analysis}, the non-linear temperature dependence of the density of water also leads to such a stability configuration. 

In this study, we continue and expand the exploration of the internally cooled system introduced in B2012 for a wide-range of parameters, with the goal to derive scalings for important physical quantities and identify different behaviors as a function of key adimensional parameters. We begin by defining the Rayleigh number for the system and performing linear stability analysis to identify the critical Rayleigh number and the most unstable mode at this critical Rayleigh number. We discuss the relevance to the model to dry atmospheric convection. We characterise the changes in various important response parameters, particularly the temperature profiles, the up-down asymmetry and the total convective heat flux as well as the convective mass flux change with changing %the various %CM: remove 
input parameters of the model. 

The article is laid out as follows. Section \ref{sec:Methods} discusses the basic equations of our model, a recap of the basic relations derived in B2012 and a description of the numerical methods. In Section \ref{sec:overview} we derive some more analytical results, define appropriate length, velocity and temperature scales to derive the non-dimensional parameters characterising the system and end with a discussion on typical values for various parameters in the atmosphere. Section \ref{sec:results} presents the linear-stability analysis for the system and the main results from our numerical simulations, following which we conclude with a discussion on the significance of the results as well as avenues for future work in Section \ref{sec:discussion}. 

\section{Methodology: Equations and numerical simulations}\label{sec:Methods}
%\ca{I'd rename : "Methodology: Equations and numerical simulations" (same order as subsections)}

\subsection{Equations}
As in B2012, we start with the incompressible Boussinesq \cite{boussinesq1903theorie} fluid equations along with the heat equation with a bulk-cooling term. We thus write the density as the sum of a reference constant density $\rho_{ref}$ and a perturbation from this reference density denoted by $\rho'$ such that $\rho=\rho_{ref}+\rho'$, with corresponding pressure $p = p_{ref}(z)+p'$ where $p_{ref}(z)$ is in hydrostatic balance with $\rho_{ref}$ so that $dp_{ref}/dz= -\rho_{ref} g$.
The equations for the total fluid velocity vector $\uu$ with vertical component $w$ and temperature fluctuations $T'$ about a reference temperature $T_{ref}$ are

\begin{gather} 
 \nab \cdot \uu = 0, \label{eq:incomp} \\
  \partial_t \uu + (\uu \cdot\nab) \uu = - (1/\rho_{ref}) \nab p' + \nu \nabla^2 \uu - \beta T' \boldsymbol{g}, \label{eq:Nav-Stokes} \\
  \partial_t T' + \uu \cdot \nab T' + \gamma w = \kappa \nabla^2 T' - R, \label{eq:Heat-eqn}
\end{gather}

where $\nu$ is the kinematic viscosity, $\beta$ is the thermal expansion coefficient, $\boldsymbol{g} = -g \hat{z}$ is the acceleration due to gravity and $\gamma = g/c_p$ is the lapse-rate of the mean stratification (or the dry adiabatic lapse-rate, see \ref{app:heateq}). The form of the equations above with the lapse-rate and the Boussinesq approximation was demonstrated first in \cite{spiegel1960boussinesq}. $R$ ($>0$) is the bulk-cooling term applied to the fluid. As discussed by B2012, the term $R$ breaks the up-down symmetry of the internally cooled system, leading to downward moving fluid ($w < 0$) occupying more than half the domain. 

Here we have assumed that the density perturbation $\rho'$ is given by $\rho'(T') =  - \rho_{ref} \beta T'$, $\rho_{ref}$ being the density of the fluid at temperature $T_{ref}$. For the Boussinesq approximation, it is only important that $T'$ remains small enough that 
$\rho'(T')/\rho_{ref} \ll 1$  %$\rho(T)/\rho_{ref} \ll 1$ %CM: I think need primes here
or equivalently, $\beta T' \ll 1$. For simplicity, we henceforth drop the primes and investigate equations $\eqref{eq:incomp}$ - \eqref{eq:Heat-eqn} without the primes. 

%The reference temperature and density does not explicitly enter the equations as pressure can be redefined as $p' \to p'  + \beta T_{ref} g z$.  

%It is important to note here that in making the Boussinesq approximation, we assume that deviations of temperature $T^{'}$ from a hydrostatic profile given by $\overbar{T}(z) = T_{bot} - \gamma z $ are small enough that relative changes in density are small, ie., $\rho^{'} / \rho_0 = -  \beta T^{'} \ll 1$. The validity of this assumption for our chosen parameters will be commented upon . 

The velocity and the temperature fields are periodic in the horizontal directions. At the bottom ($z=0$) and top ($z=L_z$) surfaces, they are given by
\begin{equation}\label{eq:fluigbound}
    \uu(z = 0) = \uu(z = L_z) = 0
\end{equation}
and 
\begin{gather}\label{eq:thermbound}
    \frac{\partial T}{\partial z} \biggr \rvert_{z = 0} = -f_0, \qquad  \frac{\partial T}{\partial z} \biggr \rvert_{z = L_z} = -f_1.
\end{gather}

\begin{figure}
    \centering
    \includegraphics[width = 0.4 \textwidth, keepaspectratio]{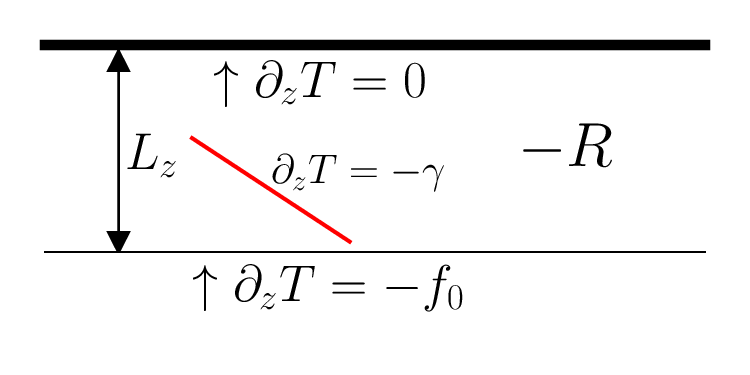}
    \caption{Schematic of the model set-up. The internal cooling R is constant everywhere in the domain and is in large-scale balance with the incoming heat-flux $f_0$ at the lower boundary while the upper boundary is insulating. The fluid has a constant temperature lapse-rate $\gamma$.}
    \label{fig:Schematic}
\end{figure}

It is noteworthy that this system is invariant under the transformation $T \to T + \delta T$ where $\delta T$ is some constant temperature since the heat equation as well as the boundary conditions contain only derivative terms of the temperature. In moist convection, this is no longer true as the partition of water into solid, liquid and gaseous phases strongly depends on the precise thermodynamic temperature. 

In addition to the fluid equations, we recall from B2012 that the large-scale thermal balance for the system, which can easily be derived by taking the domain average of eqn. \ref{eq:Heat-eqn} in steady state, is given by 
\begin{equation}\label{eq:global-heat}
    \kappa (f_0 - f_1) = R L_z. 
\end{equation}
As in B2012, we set $f_0$ to be positive and $f_1 = 0$, leading to a balance between the radiative cooling in the domain and the net heat-flux from the bottom boundary. %CM: boundaries. 
This also ensures that the average temperature of the domain remains stationary in time. Figure \ref{fig:Schematic} shows a schematic diagram of the fluid configuration. 

A steady-state solution for the equations with $\uu = 0$ everywhere leads to a temperature gradient
\begin{equation}\label{eq:stab_grad}
    \frac{dT}{dz} = - f_0 \left(1 -   \frac{z}{L_z} \right)
\end{equation}

with the corresponding temperature profile given by 

\begin{equation}\label{eq:stab_prof}
    T(z) = T_{bot} - z  f_0 \left(1 - \frac{z}{2 L_z}    \right ),
\end{equation}

where $T_{bot}$ is the temperature at $z=0$. The fluid is stably stratified where $\partial_z T > - \gamma$, which is given by $z > L_z (1 - \gamma/f_0)$. Thus we expect that when convection occurs, it occurs only up to a height $z_0$ given by 
\begin{equation}\label{eq:z0-defn}
    z_0 = L_z (1 - \gamma/f_0).
\end{equation}

It should be noted here that the equations may also be written in terms of a potential temperature $\theta \equiv T + \gamma z$. This would eliminate the adiabatic cooling term $\gamma w$ with the boundary conditions given by $\partial_z \theta = -f_0 + \gamma$ and $\partial_z \theta = \gamma$ at the lower and top surfaces respectively. Hence, the stratified system is equivalent to an unstratified fluid with fixed-flux temperature conditions at both boundaries.

\subsection{Numerical Methods}\label{sec:Numerical}
In order to understand and investigate the behaviour of the above system, we solve the system of equations \eqref{eq:incomp}-\eqref{eq:thermbound} numerically in a 2D domain box with fixed aspect ratio $2 \pi$ which is periodic in the horizontal direction. 
The various input parameters other than the extent of the domain are $\gamma$, $\kappa$, $\nu$, the product $\beta g$ and $f_0$. $f_1$ is set to $0$, $\rho_{ref}$ to unity while $R = \kappa f_0$ in accordance with the condition of thermal equilibrium (eqn. \eqref{eq:global-heat}). 
We also perform simulations for a horizontally periodic 3D domain with the same aspect ratio for a few selected parameters. The equations are solved with a python code employing the Dedalus spectral solver \cite{burns2020dedalus}, using Fourier components in the periodic horizontal directions and Chebyshev polynomial decomposition for the vertical direction. The equations are solved on a $256 \times 128$ grid for most flows, with the resolution rising to $2048 \times 1024$ for the most strongly forced flows. All 3D flows were solved on a $256 \times 256 \times 128$ grid. The code was bench-marked against the results from B2012. 

The outputs of the simulations are analysed to calculate various flow parameters only after the flow has become statistically stationary, wherein the average kinetic energy and the average temperature of the domain as well as the average temperature at the top and bottom surfaces of the domain fluctuate about a constant value. While we report only a few measurements from 3D runs, we have checked that all the below results are qualitatively true for 2D as well as 3D flows, unless otherwise mentioned. 

\section{Model Overview}\label{sec:overview}

\subsection{Exact Relations}

% \ca{This section is not a methodology, it is the beginning of results (although not all new, but still this is not our methods). I would suggest to reorganize: section 2 methodology with 2.1 equations and 2.2 numerical experiments subsections; section 3 "Overview of simulations" with 3.1 Simulations,  3.2 Exact Relations,  3.3 Non-dimensional parameters, and 3.4 "Some orders of magnitude", followed by section 4 "Results"  with 4.1 Transition to convection etc.... Then end with a last section with Discussion and Conclusiokns  }

% \ca{The subsection "3.1 Simulations" would go before this subsection "3.2 Exact Relations", to show the overview of numerical results, e.g. some snapshots (it was useful to have those from Berlengiero), with some dimensional results (e.g. show the domain mean temperature profile $T(z)$ with the $\gamma$ slope superimposed?; wb(z); A(z)...) i.e. the domain statistics/aspects related to the key points that you explain later and thus that are relevant for the remainder of the paper, to motivate the theoretical derivations.}

% \ca{Here you could add a sentence of transition " We start by recalling some of the exact relations derived in Berlengiero"...} 

In addition to the large-scale thermal balance and the steady-state temperature profile (eqns. \eqref{eq:global-heat} and \eqref{eq:stab_prof}), we also derive the z-dependent thermal balance as is standard in the case of Rayleigh-B\'enard convection and other model thermal fluid systems. Rewriting the advective term in flux-form ($\nab \cdot (\uu T)$) and integrating the horizontal statistical average of eqn. \eqref{eq:Heat-eqn} from the bottom to a height $z$ gives

\begin{equation}\label{eq:Nuss}
    \langle w T \rangle_{A,t} - \kappa \langle  \partial_z T \rangle_{A,t} = R (L_z - z), 
\end{equation}

where $\langle \cdot \rangle_{A,t}$ indicates the statistical average taken at a fixed height $z$ and we have used the fact that in the steady-state, the time-derivatives go to $0$ while horizontal derivatives of the flux term average out to zero due to the periodic boundary conditions. This relation shows that the total vertical heat transfer by convection decays linearly as a function of height. Heat is removed uniformly at each height by the bulk radiative cooling term. In case of $f_1$ not being set to $0$, there would be an addition term $\kappa f_1$ on the RHS which represents the outward heat-flux at the top of the domain. 

For our system, we also calculate the global thermal dissipation $\epsilon_T \equiv \kappa \langle |\nab T|^2 \rangle$ given by (see \ref{AppendixA})

\begin{align}\label{eq:heat_disip}
    \kappa \langle |\nab T|^2 \rangle &=  \frac{\kappa}{L_z} T_{bot} f_0 - \gamma \langle wT \rangle - R \langle T \rangle \nonumber \\
    &= R \langle T_{bot} - T \rangle - \gamma \langle wT \rangle.
\end{align}
where $\langle \cdot \rangle$ indicates the statistical average taken over the whole domain and $T_{bot}$ is the measured average temperature at $z=0$. We can see that %that %CM: remove (repeated)
the RHS value still remains invariant under the transformation $T \to T + \delta T$. The strength of the thermal gradients in the fluid increases with the average departure of the domain temperature from $T_{bot}$ and decreases when convective transfer of heat is efficient, that is $\langle wT \rangle$ is large. 

The viscous dissipation $\epsilon \equiv (\nu/2) \sum_{i,j} (\partial u_i/\partial x_j + \partial u_j/\partial x_i)^2$ is given by (see \ref{AppendixB})

\begin{equation}
    \frac{\nu}{2} \sum_{i,j} \left(\frac{\partial u_i}{\partial x_j} + \frac{\partial u_j}{\partial x_i} \right)^2 = \beta g \langle wT \rangle. 
\end{equation}

We can see that, similar to the Nusselt number in R-B convection, the response quantity $ \langle wT \rangle$ is of fundamental importance and it determines the convective heat transfer properties of the fluid system as well as the strength of the thermal and kinetic gradients. In the context of a convecting atmosphere, this value is the vertical buoyancy-flux, crucial in determining the thermodynamics of the atmosphere in shallow dry convection. In deep, moist convection too, the thermal flux in the sub-cloud layer is of crucial importance - for example it is approximately equal to the pressure-work done by the fluid and this is closely related to the production of irreversible entropy in the sub-cloud layer \cite{emanuel1996moist}. 

Analogous to the usual definition for R-B convection we define the Nusselt number $\Nus$ for the current system as 

\begin{equation}\label{eq:Nus-defn}
    \Nus = \frac{\langle w T \rangle}{\kappa T_0/L_z}
\end{equation}

where $T_0$ is a temperature scale of the system to be defined later. 
The Nusselt number here is thus a non-dimensionalised heat-flux.
The Nusselt number here compares the vertical heat-flux with the typical conductive heat flux. When the flow is conductive, we have no fluid motion and hence $\Nus = 0$. In the presence of convection, $\Nus$ has a finite value which increases with the increase in the strength of convection. 

\subsection{Non-dimensional parameters}
\subsubsection{Buoyancy Scaling}
To quantify the system and non-dimensionalise the equations, we need appropriately defined temperature, velocity and length scales. Since the dynamics of the system is set by the bulk-cooling $R$, which has SI unit dimensions of $\text{Kelvin s}^{-1}$, we can write the temperature scale $T_0$ as 

\begin{equation}
    T_0 = R t_0
\end{equation}
where $t_0$ is the appropriate time-scale, with length-scale $L_z$ giving the velocity scale $U_0 = L_z/t_0$. As is standard for thermal flows, we write $U_0$ as a convective velocity given by 

\begin{equation}
    U_0 = \sqrt{\beta g T_0 L_z}. 
\end{equation}

Combining the above equations gives finally 
\begin{gather}
    T_0 = (\beta g)^{-1/3} R^{2/3} L_z^{1/3}, \label{eq:tempscale}\\
    U_0 = (\beta g)^{1/3} R^{1/3} L_z^{2/3}, \label{eq:velscale}\\
    t_0  = (\beta g)^{-1/3} R^{-1/3} L_z^{1/3}. 
\end{gather}

Using these scales, the equations are non-dimensionalised as 

\begin{gather} 
 \widehat{\nab} \cdot \widehat{\uu} = 0, \label{eq:nondim_incomp} \\
  \partial_{\widehat{t}} \widehat{\uu} + (\widehat{\uu} \cdot \widehat{\nab}) \widehat{\uu} = - \widehat{\nab} \widehat{p} + \sqrt{\frac{\Pr}{\Ra}} \widehat{\nabla}^2 \widehat{\uu} + \widehat{T} \hat{z}, \label{eq:nondim_Nav-Stokes} \\
  \partial_{\widehat{t}} \widehat{T} + \widehat{\uu} \cdot \widehat{\nab} \widehat{T} + (\gamma/f_0) \sqrt{\Pr \Ra} \widehat{w} = \frac{1}{\sqrt{\Pr  \Ra }} \widehat{\nabla}^2 \widehat{T} - 1, \label{eq:nondim_Heat-eqn}
\end{gather}

where the hat ( $\widehat{}$ ) indicates the non-dimensionalised variable or operator obtained by dividing by the appropriate dimensional scale. 
The non-dimensional parameters are the Rayleigh-Number $\Ra$, the Prandtl number $\Pr$ and the ratio $\gamma/f_0$ between the lapse-rate and heat-flux. The Rayleigh number is given by 

\begin{equation}
    \Ra = (\beta g)^{2/3} \frac{R^{2/3} L_z^{10/3}}{\nu \kappa}. 
\end{equation}

This is equivalent to the form derived in B2012, albeit without appealing to a flux formulation. This equivalence is not surprising as the approach of B2012 (setting the time-scale of the fluid motion equal to the time-scale of radiative cooling) is identical to our approach in setting the temperature scale according to the rate of radiative cooling, which gives a non-dimensional rate of radiative cooling ($R L_z/(U_0 T_0)$) of unity. $\Ra$ can also be re-written in terms of $f_0$ using $R^{2/3} L_z^{10/3} = (\kappa f_0)^{2/3} L_z^{8/3}$ from eqn. \eqref{eq:global-heat}, which gives $\Ra \propto f_0^{2/3}/\kappa^{1/3}$. 
%We shall later see the advantage of writing down the velocity and temperature scales explicitly. 

The Prandtl number is the ratio $\nu/\kappa$ between the viscosity and the thermal conductivity. The ratio $\gamma/f_0$ sets the height $z_0$ above which the fluid is stably stratified (see Eqn. \eqref{eq:z0-defn}). It can indeed be argued that the height $z_0$ rather than $L_z$ is the appropriate length scale to define the convective time and velocity-scales, which leads to a lapse-rate dependent Rayleigh number $\Ra_\gamma$ given by 

\begin{equation}
    \Ra_\gamma = (\beta g)^{2/3} \frac{R^{2/3} z_0^{10/3}}{\nu \kappa} = \Ra \left( 1 - \frac{\gamma}{f_0} \right)^{10/3}. 
\end{equation}

In this study, we solve the system of equations in arbitrarily chosen simulations units - however, all the results presented involve only non-dimensional quantities and it is the scaling with the non-dimensional parameters that we are interested in to characterise the system. 

While several of our simulations have large differences of temperature between the lower and upper surface, thus invalidating the assumptions underlying the Boussinesq approximation, we note that this temperature difference is dictated by a combination of the temperature gradient at the lower boundary $f_0$ and the lapse-rate $\gamma$. 
The principle of dynamic similarity for flows with the same $\Ra_\gamma$ can be used to construct an equivalent flow with $f_0$ chosen to be an appropriately small value and also decreasing $\kappa$ and $\nu$ to retain the same $\Ra_\gamma$ ($\propto f_0^{2/3}/\kappa^{1/3}$).

\subsubsection{Diffusive Scaling}
It is also standard to define the velocity scale as a diffusive velocity $U_D$ given by 
\begin{equation}
    U_D = \kappa/L_z. 
\end{equation}
Defining the diffusive time-scale $t_D$ as $L_z/U_D$ and again setting $R$ to be unity, gives the non-dimensional equations
\begin{gather} 
 \widehat{\nab} \cdot \widehat{\uu} = 0, \label{eq:nondim_incomp2} \\
  \partial_{\widehat{t}} \widehat{\uu} + (\widehat{\uu} \cdot \widehat{\nab}) \widehat{\uu} = - \widehat{\nab} \widehat{p} + \Pr \widehat{\nabla}^2 \widehat{\uu} + \Pr \Ra_D \widehat{T} \hat{z}, \label{eq:nondim_Nav-Stokes2} \\
  \partial_{\widehat{t}} \widehat{T} + \widehat{\uu} \cdot \widehat{\nab} \widehat{T} + (\gamma/f_0) \widehat{w} =  \widehat{\nabla}^2 \widehat{T} - 1, \label{eq:nondim_Heat-eqn2}
\end{gather}
where $\Ra_D$ is now the diffusive Rayleigh number, related to the buoyancy Rayleigh number as
\begin{equation}
    \Ra_D = \Ra^{3/2}. 
\end{equation}
We refer the reader to \cite{goluskin2016internally} or \cite{sparrow1964thermal} for a detailed derivation of the diffusive temperature, velocity and time-scales and the above non-dimensional equations. 

We make use of the above non-dimensional equations with $\Ra_D$ for the linear stability analysis 
(\S\,\ref{subsec:stability}) %CM: Add section number
as it simplifies the algebra greatly and it has been used prominently by previous studies with fixed heat-flux boundary conditions for such stability analyses. %the same. }%CM: Rephrase
\subsection{Some Orders of Magnitude} %\ca{Change to section 3.4 "Some orders of magnitude"?}}
In the context of a typical scenario of dry convection in the tropics, convection occurs in a layer of atmosphere from the surface to cloud base level, which is $500 \text{ m}$ to $3 \text{ km}$ high, with the thermal heat flux (sensible heat flux) from the surface in the order of $10 - 100 \text{ W m}^{-2}$, radiative cooling to space of $1 - 2 \text{ K day}^{-1}$ and dry air with a typical $\beta \sim 1/300 \text{ K }^{-1} $. 

\renewcommand{\thefootnote}{\roman{footnote}}
Sensible heat flux in $\text{ W m}^{-2}$ can be converted to an equivalent thermal gradient by dividing by $(\rho \kappa c_p)$ where $\rho$ is the density and $c_p$ is the heat capacity of air. When we consider an eddy-diffusivity like value of thermal conductivity $\kappa \sim 10^{-2} \text{ m}^2 \text{s}^{-1}$, with $\rho \sim 1 \text{ kg m}^{-3}$ and $c_p \sim 10^3$ $\text{ J kg}^{-1} \text{ K}^{-1}$ we obtain $f_0 \sim \mathcal{O}(1)$. Considering the adiabatic lapse-rate to be the typical dry value of $0.01 \text{ K m}^{-1}$ and $\beta g \sim 1/30 \text{ m s}^{-2} \text{K}^{-1}$ gives $Ra_\gamma$ to be at the least of order $10^{9}$. For the true molecular value of $\nu$ and $\kappa$, the value of $\Ra_\gamma$ would be closer to $10^{15}$. It is important to note here that even for the lowest estimates of $f_0$, $\gamma$ is less than $1$\% of $f_0$ \footnote{If we consider real-world observations, it is well-known that ambient temperature measurements are recorded $2$ metres above the land surface, which is typical a few Kelvin warmer than the air at the lowest level of the atmosphere on a sunny day. Thus, the temperature gradient at the boundary here is of the order of $1$ Kelvin per metre, compared to the dry adiabatic lapse rate of $0.01 \text{ K m}^{-1}$. }. 

%\ca{\it This is a detail, but is that the value used in Berlengiero?} \lokahith{Berlengiero just jumps straight to non-dimensionalised values} 

Given that the value of $10 \text{ W m}^{-2}$ for the sensible heat-flux is a global annual average, there do exist scenarios where the sensible heat flux can be far smaller than this value. However, in these situations it is unrealistic to model the atmosphere to be locally in equilibrium with radiative cooling - instead the energy balance would be determined by large-scale, horizontal heat-fluxes or the cooling term would dominate the heat-equation. Studies of adjustment in such an out-of-equilibrium state is beyond the scope of the current study. 

In using the current system as a model for atmospheric dry-convection, it is also important to assess the validity of the Boussinesq-approximation for dry convection. For a layer of atmosphere around $1$ km thick, temperature differences $\Delta T$ between the bottom and the top of this layer are of the order of $\sim 10 \text{ K }$, as the atmosphere remains close to a dry adiabatic profile. This translates to $\beta \Delta T \approx 3 \times 10^{-2}$. Thus, we see that the Boussinesq approximation remains an excellent approximation in the simulation of dry convection. In the presence of moisture and the ensuing deep, moist convection that penetrates up to the top of the troposphere ($\sim 15$ km), the Boussinesq approximation breaks down and the vertical variation in density is significant.

% , ie., $\rho' \ll \rho_0$, which translates to $\beta |T'| \ll 1$, where $T'$ is the deviation of the fluid temperature from the reference temperature $T_{\rm{ref}}$ where the density is given by $\rho_0$. , this translates to $T' \sim 10$ Kelvin, giving $\beta T' \approx 3 \times 10^{-2}$. 

From the above discussion it is clear that the current model under investigation is valid for shallow, sub-cloud dry convection where it can be assumed that the convection and sensible heat-flux is locally in equilibrium with the cooling of the atmosphere through outgoing long-wave radiation to space. We study flows with $\Pr = 1$ and $\Ra_\gamma$ up to $10^6$. $\gamma/f_0$ is varied from $0$ to $0.75$, with the results of the simulations mainly presented for the values of $\gamma/f_0 = 0, 0.2, 0.4, 0.75$. 

%\ca{\it does that imply dT/dz is unstable to dry convection near the bottom surface? \lokahith{Here even if dT/dz > $\gamma$, it doesn't imply instability to dry convection as $\kappa$ is quite important. I guess the right answer would be it is unstable to dry convection where the system has $\Ra$ greater than the critical $\Ra$}}

\begin{table}

% \begin{minipage}{0.2\textwidth}
%     \centering
%     \begin{tabular}{|c|c|}
%     \hline
%       \textbf{Parameter} &  \textbf{Range}   \\
%     \hline
%         $\Ra_\gamma$  & ($50$, $10^6$)  \\
    
%         $\gamma/f_0$  & ($0$, $0.75$)  \\
    
%         $\Pr$  & $1$  \\
%     \hline
%     \end{tabular}
%     \captionof{table}{The range of dimensionless parameters studied}
%     \label{tab:params}
% \end{minipage}
% \quad
\begin{minipage}{0.99\textwidth}
    \centering
    \begin{tabular}{|c|c|c|c|c|c|}
    \hline
    $\gamma/f_0$ & $\Ra_\gamma$ (S) & $\Ra_c$ (S) & $\Ra_c$ (EVP) & $\Ra_c$ ($k \to 0$) &  $k_c$   \\
    \hline
    $0$    & $130.62$ & $1492.87$     & $1440.02$    & $1440$     & $0.02$   \\
    $0.2$  & $87.07$     & $2479.5$   & $2399.86$    & $2400$     & $0.02$   \\
    $0.4$  & $68.014$     & $7213.5$ & $7193.74$    & $7200$     & $0.606$  \\
    $0.5$  & $77.7$     & $21915$ & $21904.4$    & $\infty$   & $2.854$  \\
    $0.6$  & $78.695$     & $68175$ & $68139.33$   & $-7200$    & $3.804$  \\
    $0.75$ & $78.138$     & $707287.5$ & $707352.9$   & $-2880$    & $5.98$   \\
    \hline
    \end{tabular}
    \captionof{table}{Critical $\Ra_\gamma$ and diffusive Rayleigh number $\Ra_c$ for transition to convection. (S) indicates the value from 2D simulations. (EVP) indicates the numerical solution from solving the linear eigenvalue problem, ($k \to 0$) is the theoretical prediction from the long-wavelength asymptote while $k_c$ is the least stable mode from the EVP calculation. }
    \label{tab:conv_trans}
\end{minipage}
\end{table}

\begin{figure}
    \centering
    \includegraphics[width = \textwidth, keepaspectratio]{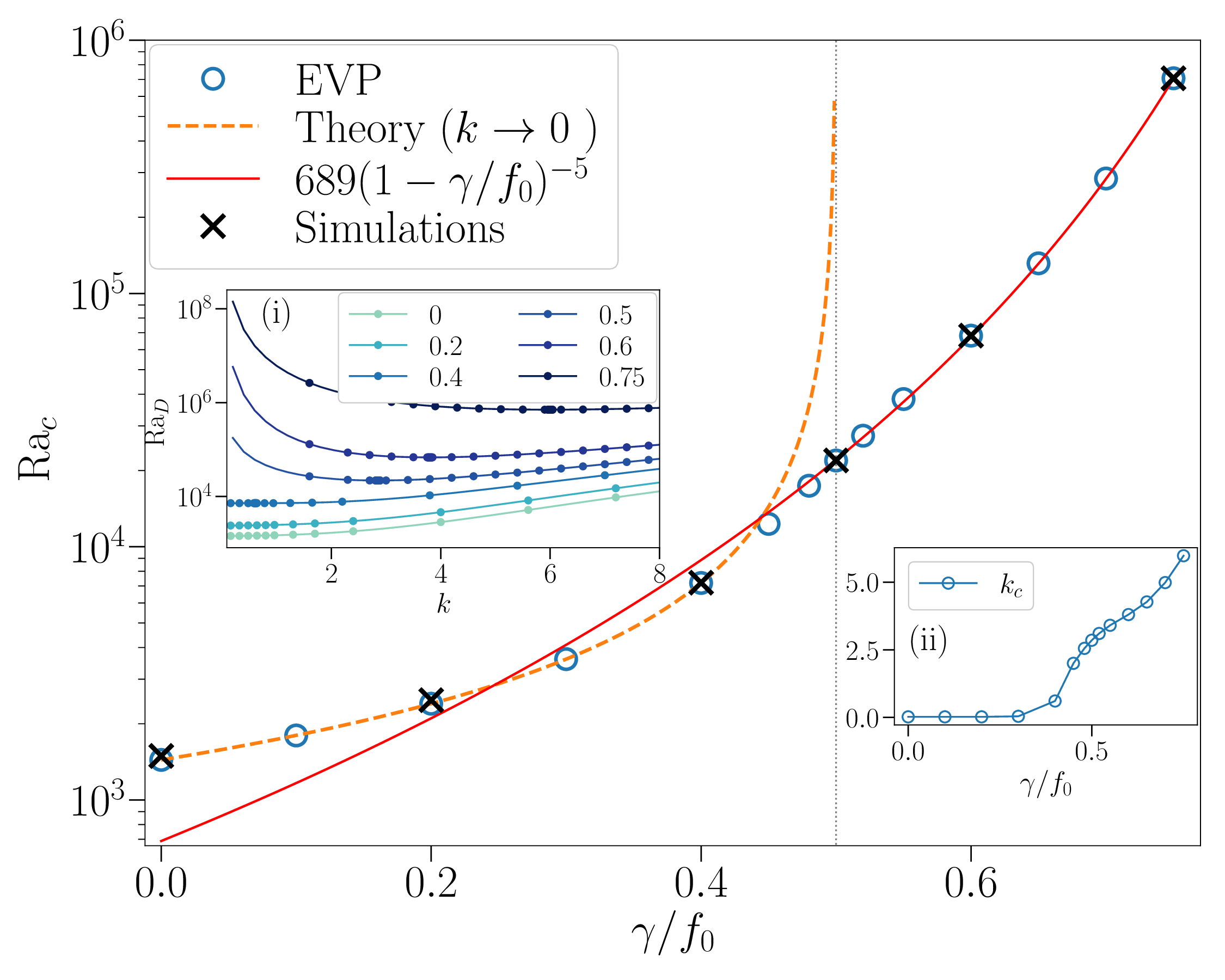}
    \caption{Variation of the critical diffusive Rayleigh number $\Ra_c$ with $\gamma/f_0$. EVP values are obtained by solving the linear eigenvalue problem numerically using the EVP command of the python package Dedalus. Theory curve shows the long-wavelength asymptote, black crosses show the values inferred from full 2D simulations while the red solid line is an empirical best-fit assuming constant $Ra_\gamma = 689^{2/3}$. Inset (i) shows the wavenumber of the most unstable mode for the linear eigenvalue problem. Inset (ii) shows the eigenvalue $Ra_D$ for varying $k$. }
    \label{fig:LinStab}
\end{figure}

\section{Results}\label{sec:results}
\begin{figure}[t]
    \centering
    \includegraphics[width = \textwidth, keepaspectratio]{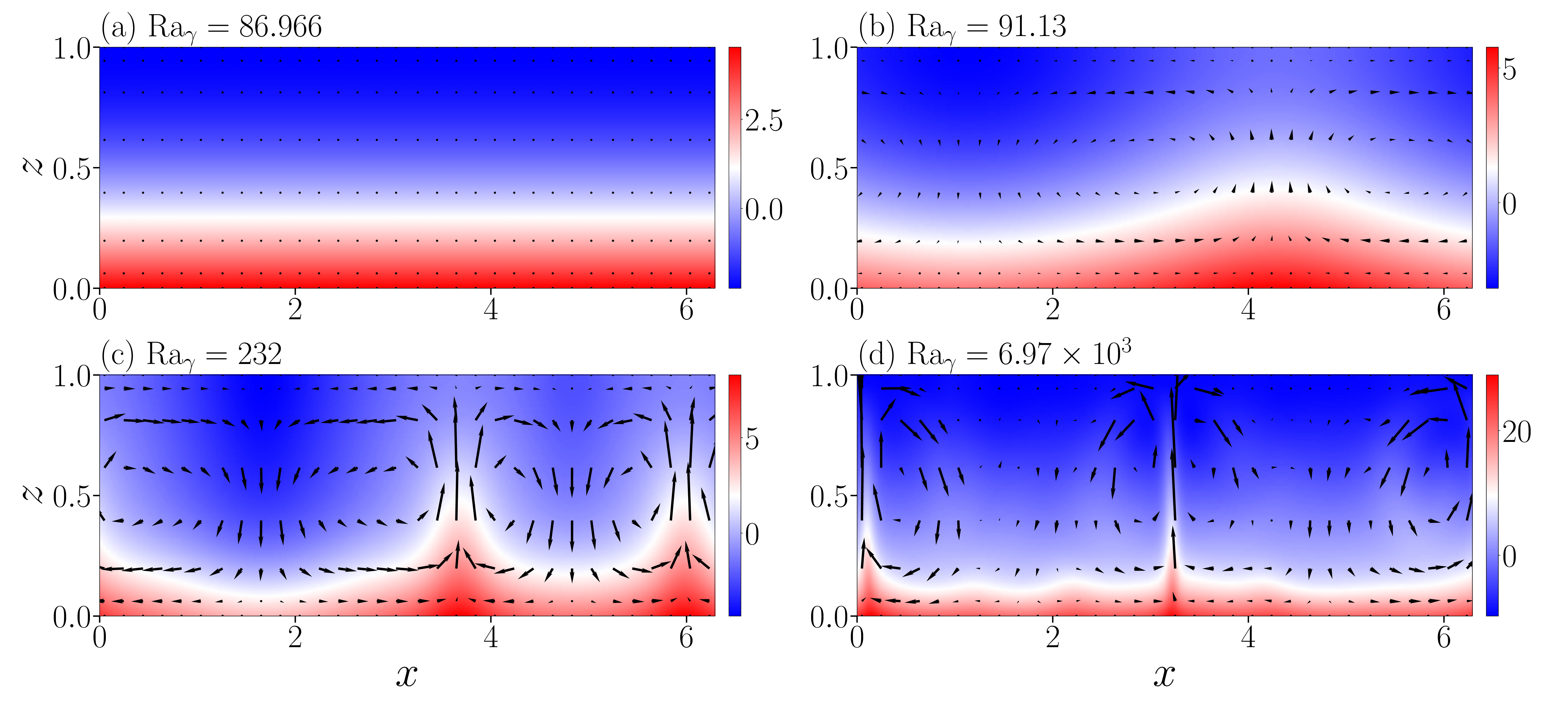}
    \caption{Instantaneous snapshots of the temperature field $T$ in the stationary regime for flows with different $\Ra_\gamma$ as indicated in the figure and $\gamma/f_0 = 0.2$. The temperature fields are divided by the respective temperature scale $T_0$ defined in Eqn. \eqref{eq:tempscale} and scaled such their domain average is $0$. Panel (a) shows a conductive state while panels (b), (c) and (d) show convective flows. The arrows represent velocity vectors with the arrow length in plot units equal to velocity magnitude divided by $U_0$, $U_0$, $2U_0$ and $5U_0$ respectively where $U_0$ is the velocity scale defined in Eqn. \eqref{eq:velscale}. }
    \label{fig:snaps}
\end{figure}

%\ca{As mentioned above, all the technical discussion of the simulations should go in the Methods section 2.2 (including table 2). 
%On the other hand, table 1 is a result, and would fit nicely in the 4.1 section Transition to convection. }
\subsection{Linear Stability Analysis}\label{subsec:stability}
The procedure to find the critical Rayleigh number for a convective instability in a thermal fluid has been well established \cite{rayleigh1916,chandrasekhar2013hydrodynamic}. We consider the conductive state with $\uu=0$ and the temperature field varying only in the vertical according to eqn. \eqref{eq:stab_prof}. Now we consider a small perturbation in velocity and temperature given by $\boldsymbol{U} = (U,W)$ and $H$ respectively. The non-dimensional equations of motion then become \cite{goluskin2015internally} 

\begin{gather} 
 \nab \cdot \boldsymbol{U} = 0, \label{eq:nondim_incomp3} \\
  \partial_{t} \boldsymbol{U}  = - \nab P + \Pr \nabla^2 \boldsymbol{U} + \Pr \Ra_D H \hat{z}, \label{eq:nondim_Nav-Stokes3} \\
  \partial_{t} H + (\gamma/f_0) W - (1-z) W =  \nabla^2 H. \label{eq:nondim_Heat-eqn3}
\end{gather}
where $P$ is the pressure perturbation with the stable temperature gradient absorbed.  Here, the non-linear advective terms are neglected and the stable temperature profile enters the equations only via its derivative $(1-z)$. The boundary conditions on $W$ and $H$ at the top and bottom surfaces are given by 
\begin{equation}
    W  \rvert_{z = 0,1} = \partial_z W  \rvert_{z = 0,1} =0; \partial_z H  \rvert_{z = 0,1} = 0. \label{eq:pert_bounds}
\end{equation}
The boundary conditions $W$ arise from the no-slip velocity boundary condition (eqn. \eqref{eq:fluigbound}) and the incompressibility condition respectively while the condition on $H$ arises from the fact that for the fixed heat-flux, $\partial_z T$ at the boundaries is fixed and constrained by the large-scale heat-balance, meaning that there can be no local perturbations from this fixed flux value. 

The curl is applied twice ($\nab \times \nab \times$) to the momentum equation eqn. \eqref{eq:nondim_Nav-Stokes3}, the vertical component of which gives 
\begin{equation}
    %\frac{1}{\Pr} \partial_t w = \nabla^4 w + Ra_D (\partial_x^2) H, %CM: PLease check but I think the equation needs to be fixed, here is what I get:
    \frac{1}{\Pr} \partial_t (\partial_z^2) W = \nabla^4 W + Ra_D (\partial_x^2) H, 
\end{equation}
while the heat-equation is simplified to 
\begin{equation}
    \partial_{t} H =  \nabla^2 H + (1 - \gamma/f_0 -z) W. 
\end{equation}
In looking for marginally stable states that do not vary in time ($\partial_t \to 0$), the problem reduces to 
\begin{gather}
    \nabla^4 W = - \Ra_D \partial_x^2  H, \label{eq:Mom_stationary_pert}\\
    \nabla^2 H = - (1 - \gamma/f_0 - z) W. \label{eq:Heat_stationary_pert}
\end{gather}
The critical Rayleigh number $\Ra_c$ for which the system is unstable to %non-linear %CM: Why nonlinear, we linearized around the reference state no?  
perturbations is given by the smallest $\Ra_D$ for which the system of equations \eqref{eq:Mom_stationary_pert} - \eqref{eq:Heat_stationary_pert} has a real, non-zero solution. This eigenvalue problem can be decomposed into Fourier modes by considering the pertubations to have the form 
\begin{gather}
    W = \mathcal{W}(z) e^{ikx},\\
    H = \mathcal{H}(z) e^{ikx},
\end{gather}
where $k$ is the horizontal wavenumber of the disturbance. This leads to 
\begin{gather}
    \left(\frac{d^2}{dz^2} - k^2\right)^2 \mathcal{W} = \frac{d^4 \mathcal{W}}{dz^4}  - 2 k^2 \frac{d^2 \mathcal{W}}{dz^2}  + k^4 \mathcal{W} - = k^2 \Ra_D  \mathcal{H}, \label{eq:Mom_stationary_Fourier}\\
    \frac{d^2 \mathcal{H}}{dz^2} - k^2 \mathcal{H} = - (1 - \gamma/f_0 - z) \mathcal{W}. \label{eq:Heat_stationary_Fourier}
\end{gather}
For the equivalent case of uniformly heated convection with a constant outward heat-flux at the top, the solution to the above eigenvalue problem 
without stratification ($\gamma/f_0=0$) %CM: add that unstratified
is found by considering the long-wavelength asymptote, ie., the limit $k \to 0$. 
%CM: Add transition and subsubsections titles for clarity? 
We will first derive the $k \to 0$ critical Rayleigh number and will see that it indeed yields the most unstable mode for weak stratification. We will then extend the results to finite stratification, which instead occurs at non zero wavenumber $k_c \ne 0$. 
\subsubsection{Long-wavelength solution for weak stratification} %CM: add subsubsection title
Focusing here on the $k = 0$ limit, %Noting that $k = 0$ is a singular limit, %CM: Rephrase
the expansion in $k^2$ of the form

\begin{gather}
    \mathcal{W} = \mathcal{W}_0 + k^2 \mathcal{W}_2 + k^4 \mathcal{W}_4 + \dots, \\
    \mathcal{H} = \mathcal{H}_0 + k^2 \mathcal{H}_2 + k^4 \mathcal{H}_4 + \dots,
\end{gather}
yields
\begin{equation}
    \mathcal{W}_0 = 0; \quad \mathcal{H}_0 = 1; \quad \frac{d^4 \mathcal{W}_2}{dz^4} = \Ra_c \mathcal{H}_0; \quad \frac{d^2 \mathcal{H}_2}{dz^2}  - \mathcal{H}_0 = - (1 - \gamma/f_0 -z) \mathcal{W}_2. \label{eq:k2-terms}
\end{equation}
where $\Ra_c$ is the critical $Ra_D$ for transition to convection. Since $\mathcal{H}_0$ is a constant and the 4th-derivative of % and %CM: remove extra "and"
$\mathcal{W}_2$ is a constant, $\mathcal{W}_2$ must be given by  
\begin{equation}
    \mathcal{W}_2 = Ra_c P^{(4)}(z), \label{eq:W2eqn}
\end{equation}
where $P^{(4)}(z)$ is a 4th-order polynomial in $z$ with the coefficients to be determined from the boundary conditions given in eqns. \eqref{eq:pert_bounds}. 
Requiring that each of the $\mathcal{W}_i, \mathcal{H}_i$ follow the same boundary conditions as $W$ and $H$ respectively, yields 
\begin{equation}
    P^{(4)}(z) = \frac{1}{24} (z^4 - 2z^3 + z^2). \label{eq:P4eqn}
\end{equation}
Integrating the right-most equation of %eqn. %CM: remove (strange to refer to right-most equation of an equation) or maybe replace with eqns. ? 
\eqref{eq:k2-terms} from the limits $0$ to $1$ and noting that $\mathcal{H}_0 = 1$, $d_z \mathcal{H} \rvert_{z=0,1} = 0$ leads to 
\begin{equation}
    1 = \frac{Ra_c}{24} \int_0^1 (1 - \gamma/f_0 -z) (z^4 - 2z^3 + z^2) dz
\end{equation}
after substituting for $\mathcal{W}_2$ from eqns. \eqref{eq:W2eqn} and \eqref{eq:P4eqn}. Finally, we derive the expression for the critical diffusive Rayleigh number as 
\begin{equation}
    \Ra_c = \frac{24}{\int_0^1 (1 - \gamma/f_0 -z) (z^4 - 2z^3 + z^2) dz} = \frac{1440}{1 - 2\gamma/f_0}. \label{eq:smallkRac}
\end{equation}
We note here that all of the above analysis has been performed by Goluskin \cite{goluskin2015internally,goluskin2016internally} with $\gamma = 0$. While Goluskin considered an internally heated system with $\partial_z T > 0$ for the conductive profile, leading to $- z W$ instead of $-(1-z) W$ in the LHS of eqn. \eqref{eq:nondim_Heat-eqn3}, this still leads to the same 4th order polynomial form for $\mathcal{W}_2$ and the final integral yields $\Ra_c = 1440$.

\subsubsection{Non-zero wavenumber solution for stronger stratification} %CM: Add subsubsection
For $\gamma/f_0 > 0$, $\Ra_c$ from eqn. \eqref{eq:smallkRac} increases, approaching $\infty$ for $\gamma/f_0 \to 0.5$. When $\gamma/f_0 > 0.5$, $\Ra_c$ is negative. Here, the small wavenumber (or long-wavelength) approximation is no longer valid as the mode of least stability doesn't approach the $k=0$ mode. The mode of least stability is the value of wavenumber $k$ for which the linear eigenvalue problem in $Ra_D$ yields the smallest, real solution.  We denote this wavenumber as $k_c$. 

We use two approaches to infer the critical $Ra_D$ for transition to convection in our system for varying $\gamma/f_0$. In the first approach, we use full 2D simulations as described in section \ref{sec:Numerical} and examine whether the flow shows any convective motion or remains conductive. The flow is initialised with a stable temperature profile with spatially varying random Gaussian noise. This leads to an initial spike in kinetic energy following which, the kinetic energy either falls to $0$ (ie., extremely small values below $10^{-25}$ in the simulation units) or remains non-zero accompanied by a non-zero value for the convective heat-flux. 

Secondly, we directly solve the linear eigenvalue problem described by eqns. \eqref{eq:Mom_stationary_Fourier} - \eqref{eq:Heat_stationary_Fourier}. This is achieved using the EVP (Eigenvalue Problem) class in the Dedalus Package, numerical details of which may be found in  \cite{burns2020dedalus}. Formulations of such linear stability problem leading to similar eigenvalue equations have been considered previously for various problems. The most generalised formulation was provided in \cite{sparrow1964thermal} for convection with a constant internal heat-source or sink and any combination of fixed-temperature or fixed-flux thermal boundary conditions at the top and bottom walls. The eigensystem was explicitly solved numerically only for the case of fixed temperature boundary conditions with an internal heat source. Other studies considered internally cooled system with fixed temperature at the lower boundary \cite{hartlep2018convection}, internally heated convection with a varied combination of thermal boundary conditions \cite{goluskin2015internally,goluskin2016internally}, convection heated with a constant flux from below with no internal heat-source \cite{choi1985stability}, compressible convection \cite{depassier1981large} and convection in the water-ice system where density varies non-linearly with temperature \cite{roberts1985analysis}. To the best of our knowledge, the solution to the linear stability eigenvalue problem for the current system has not been reported elsewhere in the literature, though the methods have been discussed.  

The results are summarised in Figure \ref{fig:LinStab} and Table \ref{tab:conv_trans}. The main figure shows that the long-wavelength asymptote (orange, dashed line) for the critical Rayleigh number for transition to convection is an excellent estimate up to $\gamma/f_0 \sim 0.4$ inferred through both, simulations (black crosses) and through solving the eigenvalue problem (empty blue circles) numerically. As $\gamma/f_0$ approaches the value of $0.5$, indicated by the dotted gray vertical line, $\Ra_c$ 
from the simulations and the EVP theoretical solutions %CM: add 
departs from this prediction. Instead, the values  for larger $\gamma/f_0$ lie on the $(1-\gamma/f_0)^{-5}$ curve. %CM: is that curve derived the EVP theoretical solution? Or is it empirically determined? Maybe say here.
This corresponds to a constant $\Ra_\gamma \approx 78$. 

The insets of the figure show the behaviour of the eigenvalue $\Ra_D$ for perturbations with varying wave-number $k$ inferred from the EVP method. Inset (i) plots the eigenvalue $\Ra_D$ as a function of $k$ for different values of $\gamma/f_0$. When $\gamma = 0$, the eigenvalue is a monotonically increasing function of $k$, consistent with the long wavelength instability. %CM: add
Upon increasing $\gamma/f_0$, the shape of the curve changes, with the value of $\Ra_D$ decreasing before reaching a minimum value and then increasing for larger $k$. For example, for $\gamma/f_0 = 0.4$, $k_c = 0.606$, with $Ra_c = 7193.74$, marginally smaller than the theoretical value of $\Ra_c = 7200$. Thus, the long-wavelength asymptote remains an excellent approximation even when only the lower $60\%$ of domain height is unstably stratified. 

As $\gamma/f_0$ is increased further, the value of $k_c$ starts to significantly increase and the local minima in the curve starts to become more pronounced. This transition is apparent from comparing the $\gamma/f_0 = 0.4$ and $\gamma/f_0 = 0.5$ curves in inset (i). Now, the smallest wavenumber is no longer the most unstable mode, with $k_c$ increasing to nearly $6$ for $\gamma/f_0 = 0.75$. 

Table \ref{tab:conv_trans} lists the numerical values for the critical $\Ra_\gamma$ and $\Ra_D$ found using full 2D fluid simulations. These are compared with the long-wavelength asymptote, the numerical EVP solution as well as the $k_c$ found from the EVP procedure. These are the same as the plotted values, tabulated for more clarity. 

\subsection{Transition to convection}

\begin{figure}[t]
    \centering
    \includegraphics[width = \textwidth, keepaspectratio]{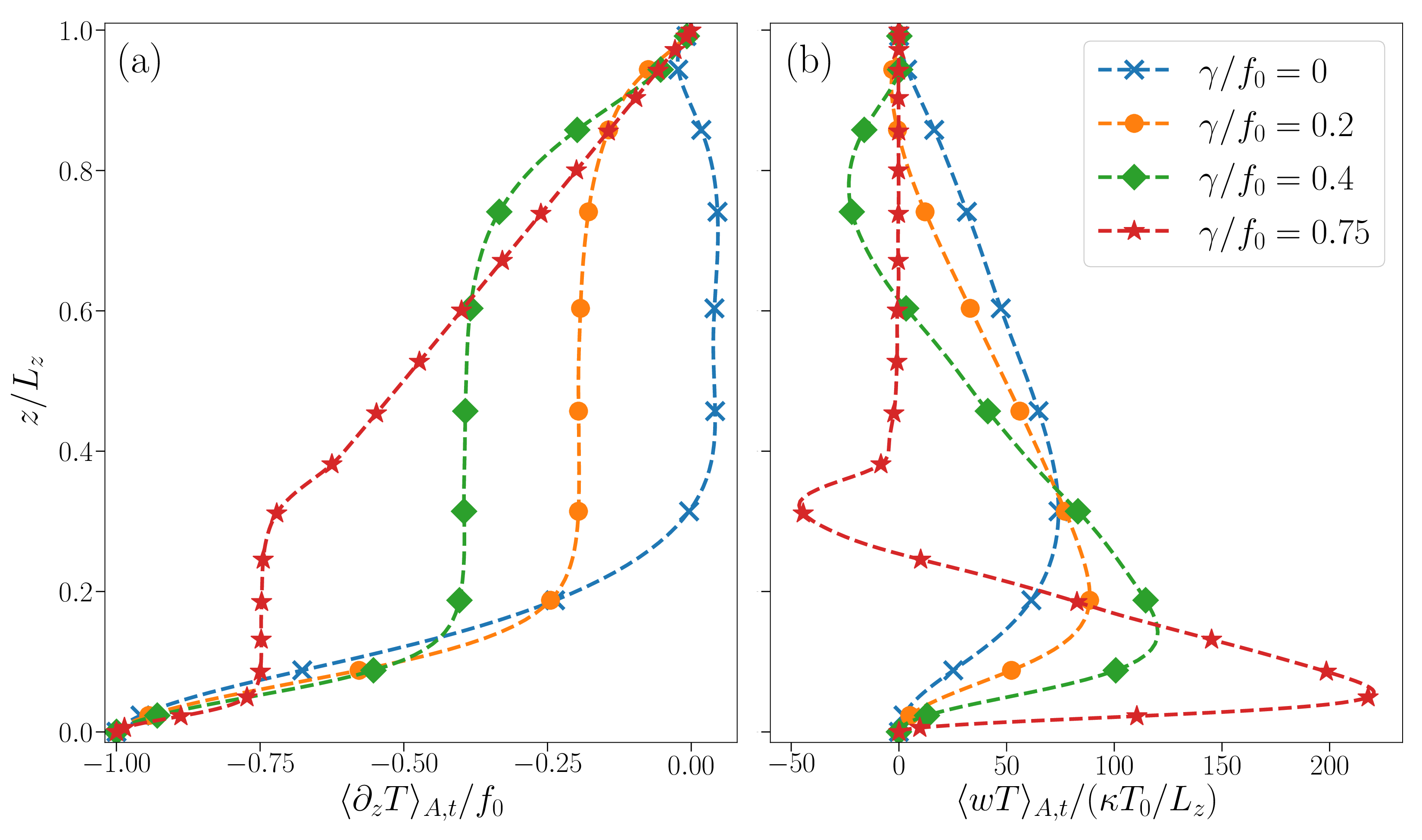}
    \caption{(a) Average vertical profiles of $\partial_z T$ divided by $f_0$. (b) Average vertical profiles of $w T$ divided by $\kappa T_0/L_z$. Both panels share the same legend, with the flows having $\Ra_\gamma \sim 10^{4}$ and the averages are statistical, height-wise averages for each flow.     }
    \label{fig:vert_profs}
\end{figure}

When $\Ra_\gamma$ is below the critical value, the fluid remains motionless. When $\Ra_\gamma$ (or equivalently, $f_0$) is increased, the flow becomes convective with convective rolls that extend up to height $z_0$. Figure \ref{fig:snaps} shows instantaneous snapshots of the linearly scaled, non dimensionalised temperature field for 2D flows with $\gamma/f_0 = 0.2$. When the flow is stable (see panel (a)), the fluid is held motionless by viscosity and thermal dissipation and the temperature field is perfectly homogeneous in the horizontal direction. When $f_0$ is increased slightly, as in panel (b), the flow shows a pattern of large, space-filling convective rolls. Larger values of $f_0$ shown in the lower panels show intense and clearly defined rising convective plumes, where the hot, rising plumes have a large positive height-wise temperature anomaly, while the gently subsiding regions outside these plumes have a smaller negative height-wise temperature anomaly. In panel (d), the hot plume is the narrowest with the highest velocities (seen by length of the arrows) strongly concentrated in the rising plumes. The reader should note that the arrows representing the velocity fields are scaled by a different value in each panel only to ensure clarity in the figure - arrow lengths must not be used to compare the velocity between different panels. 

% With $\gamma = 0$, it is found that for very small values of $f_0$ (or equivalently, small values of $R$), the flow remains motionless ($\uu = \boldsymbol{0}$) and assumes a temperature profile given by Eqn. \eqref{eq:stab_prof}. When $f_0$ is increased beyond a critical value, the flow becomes convective with convective rolls that extend up to height $L_z$ since $z_0 = L_z$ when $\gamma = 0$. For finite values of $\gamma$, the flow remains stable for larger values of $f_0$ than for the $\gamma = 0$ case. On increasing $f_0$ beyond the critical value, the flow becomes convective where the convective rolls extend up to height $z_0$. 
% For the case of $\gamma = f_0$, the flow is always stable as in this case no heat-transfer occurs from the lower surface to the fluid. We find numerically that the critical $\Ra_\gamma$ for the triggering of convection depends weakly on the ratio $\gamma/f_0$. The numerically estimated values using 2D simulations are given in Table \ref{tab:conv_trans}. The goal of the current study is to characterise the behaviour of the fully convective regime rather than a detailed study of the fluid instability. We note however that the estimation of the critical $\Ra_\gamma$ for different values of $\gamma$ should be analytically tractable through the analysis into normal modes of a small perturbation on the steady-state solution as is usually done for classical Rayleigh-B\'enard convection (for example see \cite{chandrasekhar2013hydrodynamic}) and other instabilities. 

\subsection{Flow structures and Profiles}
In the convective regime, intense hot plumes arise from the lower surface along with broad regions of cold subsidence. There exists a small diffusive boundary layer near the lower surface underneath the intense hot plumes. Here, the vertical gradient of the temperature field ($\partial_z T$) quickly goes from $-f_0$ to $-\gamma$ as seen in Figure \ref{fig:vert_profs}(a). Above this layer lies a convective bulk region where the effective mixing of the fluid ensures $\partial_z T \approx -\gamma$ up to a height of approximately $z_0$. For small $\gamma$, this corresponds to the top of the domain and thus there is no conducting region close to the upper boundary. Above $z_0$, $\partial_z T$ goes to $0$ at $z=L_z$ in a linear fashion, which corresponds to the stable layer above the convective layer. 

\begin{figure}
    \centering
    \includegraphics[width = 0.9\textwidth, keepaspectratio]{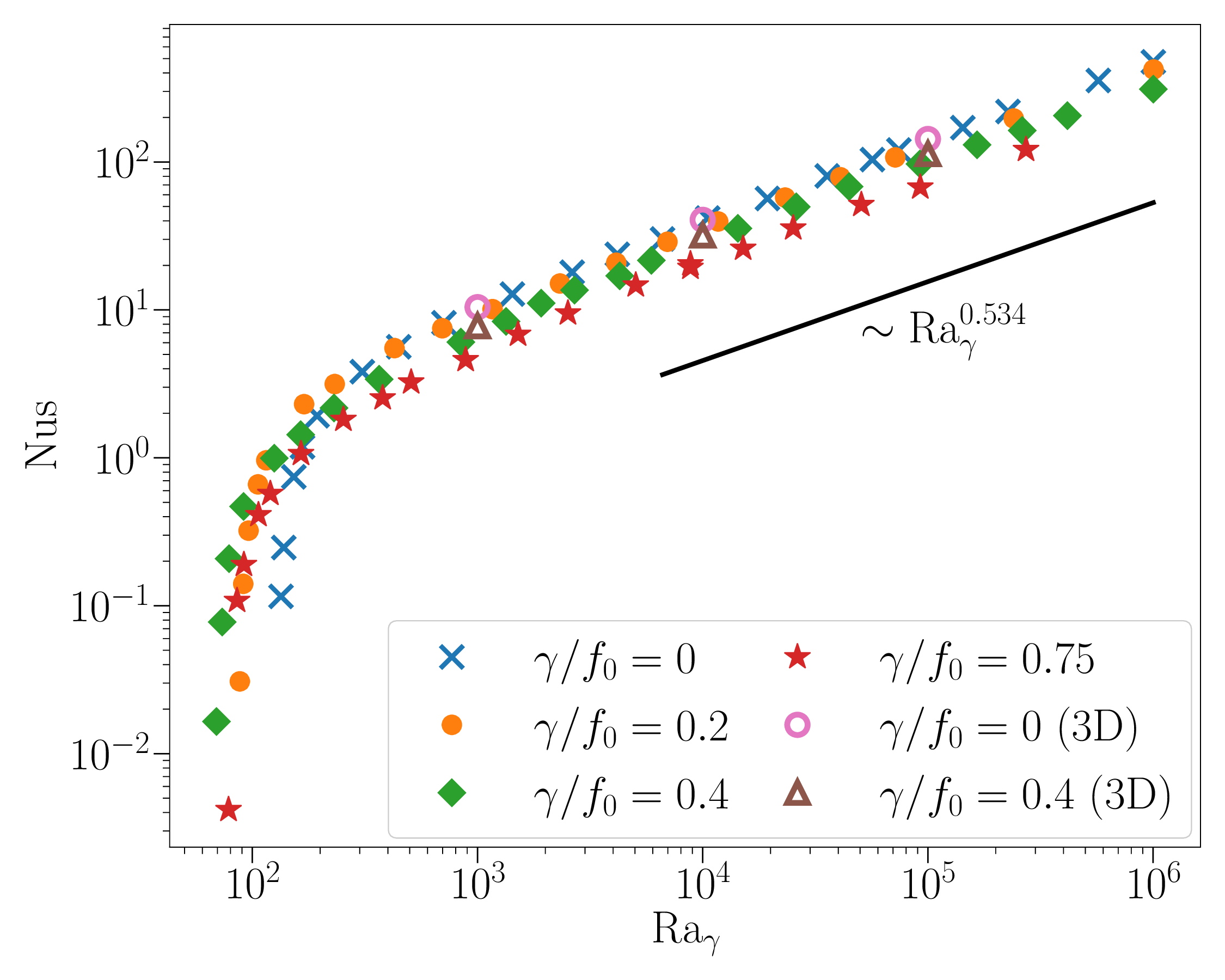}
    \caption{Nusselt number of the flow as a function of $\Ra_{\gamma}$ for $4$ different values of $\gamma/f_0$ in 2D and $2$ different values in 3D.} 
    %\caption{Scaling of the Nusselt number with $\Ra_{\gamma}$ for different values of $\gamma/f_0$. \lokahith{More points to be filled in, taking Ra upto 10^7. } } 
    \label{fig:Nuss-glob}
\end{figure}

Panel (b) of Figure \ref{fig:vert_profs} shows the time-averaged convective flux divided by the product of a diffusive velocity scale $\kappa/L_z$ and the temperature scale $T_0$ - that is the height-wise Nusselt number. The convective flux attains a maxima at $z$ roughly corresponding to the end of the lower thermal boundary layer. Above this layer and up to $z=z_0$, ie., in the convective region where $\partial_z T$ remains nearly constant, the flux decreases linearly consistent with eqn. \eqref{eq:Nuss}. 

For larger values of $\gamma$, there exists a small layer of fluid close to $z \gtrsim z_0$ where the flux is negative - this can be explained by the formation of cold patches of fluid above the rising thermal plumes. The fast rising parcels of fluid are cooled rapidly due to the lapse-rate term, leading to a region where the fluid is colder than the horizontal average  while the fluid is still rising ($w>0$), leading to a negative flux. Above this, the fluid is no longer convecting, leading to a flux close to $0$. Thus, in the presence of a large lapse-rate, the domain-averaged convective heat-flux ($\Nus$) is partially decreased by the strong cooling of fast-rising parcels of fluid. 

\subsection{The Nusselt Number}

%\ca{Here again, we need a sentence of transition to help the reader follow the logic of the paper. e.g. "because it characterizes the behavior (...) of the system, the Nusselt number is particularly important..."}

The convective heat-flux characterises the behaviour of the flow to a large degree, as it determines the strength of the velocity and temperature gradients and also plays an import role in understanding the behaviour of the fluid in the viscous boundary layers. The non dimensionalised heat-flux, the Nusselt number (defined in Eqn. \eqref{eq:Nus-defn}) is thus particularly important in understanding the convective behaviour of the system. 

The Nusselt number is identically $0$ for the conductive state as the fluid is not in motion. In the convective regime, $\Nus$ increases with increasing $\Ra_\gamma$ as we see in figure \ref{fig:Nuss-glob}, which shows the scaling of the Nusselt number for changing $\Ra_\gamma$ and for various values of $\gamma/f_0$. For flows with identical $\Ra_\gamma$, $\Nus$ is the largest for the cases with $\gamma = 0$. This is due to the fact that for flows without an imposed lapse rate, the convection is penetrative ($z_0 = L_z$) and the hot plumes reach the top of the domain. In the case of flows with a finite lapse rate, the region of the flow above $z_0$ contributes negatively to the total convective heat flux of the system, as already discussed in the previous section and seen in Figure \ref{fig:vert_profs}(b). We note that this contribution is quite small, given that the values are nearly identical for $\gamma/f_0$ ranging from $0$ to $0.75$. This also indicates that $\Ra_\gamma$ is a well-chosen parameter to characterise the large-scale behaviour of this model system. 

The magnitude and scaling of the Nusselt number as a function of $\Ra_\gamma$ for 3D flows is identical to the 2D flows. This indicates that in making estimates of dry convective heat-fluxes with given boundary conditions, it is sufficient to simulate a 2D domain. 

\subsection{Mass-flux, Updraft Velocities and Up-down Asymmetry}

Along with the convective heat-flux, the convective mass-flux also remains an important parameter to be estimated in thermal flows, particularly in the dry atmosphere. The mass flux at a given height is calculated as the product of the volume fraction $V_f$ of the domain  which is rising ($w>0$) and the average vertical velocity $U_u$ within these rising regions. The total convective mass flux is the vertical average of the mass-flux at each height written as $\langle V_f U_u \rangle$. 

\begin{figure}
    \centering
    \includegraphics[width = \textwidth, keepaspectratio]{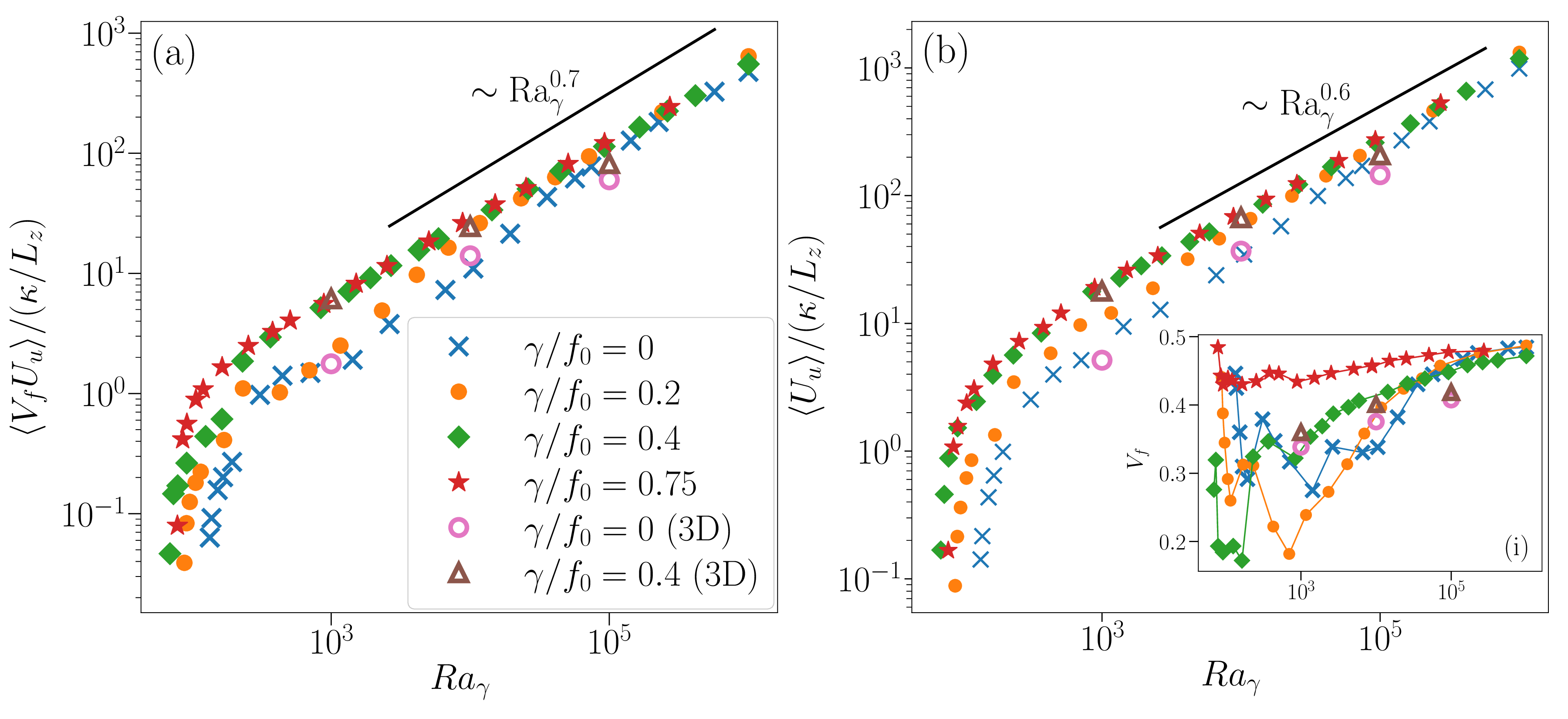}
    \caption{(a) Domain-averaged convective mass flux plotted against $\Ra_\gamma$ for $4$ different values of $\gamma/f_0$ in 2D and $2$ values of $\gamma/f_0$ in 3D. The mass flux is non-dimensionalised by dividing it by the diffusive velocity scale $\kappa/L_z$. (b) Average velocity $U_u$ in regions of the flow where $w>0$ divided by the diffusive velocity scale as a function of $\Ra_\gamma$ for the same parameters as panel (a). Inset (i) shows the volume fraction $V_f$ of the domain occupied by updrafts ($w>0$). } 
    \label{fig:Vf_Mc}
\end{figure}

The variation of the convective mass-flux with $\Ra_\gamma$ is shown in panel (a) of figure \ref{fig:Vf_Mc}. We see that the mass flux close to the transition to convection shows large variation in magnitude depending on the value of $\gamma/f_0$, with the mass flux being greater for the flows with greater $\gamma/f_0$. For large enough $Ra_\gamma$, all values of mass-flux converge to a single line on the log-log plot, which corresponds to a power law dependence on $\Ra_\gamma$ with a fixed exponent.
The updraft velocities $U_u$ and the volume fraction $V_f$ are shown in panel (b) and inset (i) respectively. For larger values of $\Ra_\gamma$ the values of $U_u$ also converge to a single exponent independent of $\gamma/f_0$ while $V_f$ converges to a single value $\sim 0.48$. 

The variation of $V_f$ with increasing $\Ra_\gamma$ with can be explained from the behavior seen in Figure \ref{fig:snaps} - when the flow is convective and laminar (as is the case for small $\Ra_\gamma$ close to the critical value for transition) the only upward moving regions occur in the vicinity of the rising plume, and the entire domain is filled by a single, large convective roll. The rest of the domain is uniformly, weakly subsiding. This behaviour can be observed for example on closer examination of panel (b) in Figure \ref{fig:snaps}, where a part of domain extending from approximately $x = 3$ to $x = 5.5$ is a clearly delineated upward rising region and the rest of the domain is subsiding with a much smaller velocity to compensate for the rising mass in the rising plume. For larger $\Ra_\gamma$ flows which are more turbulent and have greater kinetic energy, the flow contains multiple plumes and convective rolls with rapid point-wise temporal fluctuations in velocity. For example, in panel (d) of the same figure, while the region between $x \approx 1$ to $x \approx 2.5$ is subsiding on average, some grid points individually have $w>0$ within this region. For the largest values of $\Ra_\gamma$, the flows are highly turbulent and energetic, leading to a $V_f$ value close to half. 

The asymmetry between up and down that results from the $-R$ term is important only in the convective region of the flow. The stable upper layer in flows with non-zero $\gamma$ are locally in conductive equilibrium and fluid motion here is only due to the transport of mass and momentum from the convective region. These regions do not show an up-down asymmetry. Thus, regions with a thicker stable layer (larger $\gamma/f_0$) are overall more symmetric and the domain average $V_f$ is closer to half, independent of $\Ra_\gamma$.
%In flows with larger $\gamma/f_0$ (thicker stable layer), the contribution of the stable region to the domain average  thus have a domain average $V_f$ closer to half, independent of $\Ra_\gamma$. 

The mass flux and the updraft velocity scales nearly identically in 2D as well as 3D, though in the 3D case $V_f$ converges to a value close to half for much larger $\Ra_\gamma$ (if we assume that the scaling seen in Figure \ref{fig:Nuss-glob}(i) holds for larger $\Ra_\gamma$). Simulations for 3D flows with $\Ra_\gamma > 10^5$ were not performed due to the requirement of large computing resources and are beyond the scope of this study. 

\begin{figure}
    \centering
    \includegraphics[width = \textwidth, keepaspectratio]{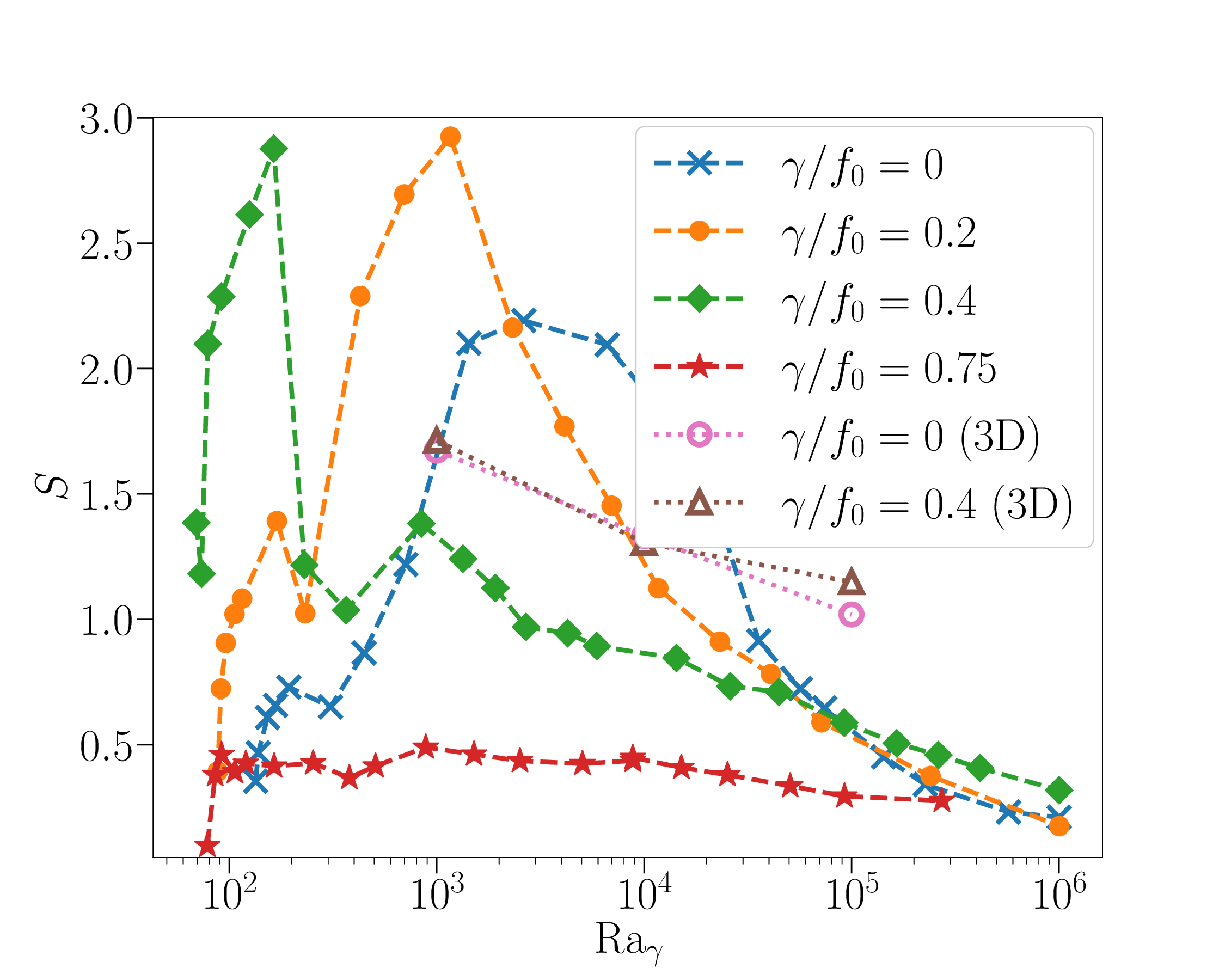}
    \caption{Skewness $S$ of the vertical velocity for flows with varying $\gamma/f_0$ as a function of $\Ra_\gamma$ for 2D and 3D simulations. } 
    \label{fig:skews}
\end{figure}

While the volume fraction $V_f$ gives an insight into the degree of asymmetry in the flow, it does not capture the relative magnitude of the velocity of rising and subsiding regions, thus does not fully quantify the asymmetry. Another measure often used (including in B2012) is the skewness measure of the vertical velocity field. The skewness $S$ is defined as 

\begin{equation}
    S = \frac{\langle w^3 \rangle}{(\langle w^2\rangle)^{3/2}},
\end{equation}

where we have used the fact that $\langle w \rangle =0$. A positive skew implies a left-peaked distribution, where the tail on the positive end is thicker while the median $w$ is negative. Figure \ref{fig:skews} shows $S$ as a function of $\Ra_\gamma$ for various values of $\gamma/f_0$. $S$ follows a trend which is close to the inverse of $V_f$ (see figure \ref{fig:Vf_Mc}(i)), where the value of $S$ peaks when the value of $V_f$ is at a minima. Similar to $V_f$, the skewness also converges to a value which is between $0$ and $0.5$ for large $\Ra_\gamma$, consistent with the aforementioned small-degree of asymmetry. 

It is the $-R$ term that breaks the up-down asymmetry - however we see that for greater $R$, the flow becomes less asymmetric. Whereas the flows with the largest degree of asymmetry have $R$ just large enough to overcome the stabilisation by viscous forces. We note here that the key factor determining the asymmetry is the relative importance of the bulk-cooling term. 
Considering the non-dimensionalised heat-equation, eqn. \eqref{eq:nondim_Heat-eqn}, the LHS value is large when the typical magnitude of fluctuations in $\widehat{T}$ and $\widehat{\uu}$ are large. This is the case when $\Ra$ is large. For cases of smaller and intermediate values of $\Ra$, the magnitude of fluctuations in $\widehat{T}$ and $\widehat{\uu}$ are small, so the cooling term is relatively more important in the dynamics. 

\section{Discussion and Conclusions}\label{sec:discussion}
We have presented results from 2D as well as 3D simulations of an uniformly cooled thermal fluid system in the presence of gravity and a lapse rate with a uniform heat flux at the lower surface and an adiabatic (no heat flux) boundary condition for the top surface. The fluid is assumed to be incompressible and we work in the regime of the Boussinesq approximation, which is a fair approximation for the sub-cloud layer of the earth's atmosphere, where the atmosphere is either dry or moist-unsaturated. The constant, bulk cooling term is set such that it balances the incoming heat flux at the lower surface, leading to a situation where there is balance between a constant cooling of the domain and a heating from below, which is analogous to the atmosphere which is heated by the earth surface and constantly loses heat to space through longwave radiation. The current work is a direct extension of previous work by Berlengiero et. al. reported in B2012. 

The equations are non-dimensionalised using a temperature and velocity scale based on the magnitude of the diabatic cooling term. For the length-scale, we choose the height up to which the domain is unstable to dry convection $z_0$, where $z_0$ depends on the ratio between the lapse-rate $\gamma$ and the uniform flux at the lower surface $f_0$. This leads to the lapse-rate dependent Rayleigh number denoted $\Ra_\gamma$ which characterises the system well. The equations are also non-dimensionalised using a diffusive velocity scaling as has been done in previous studies to give the diffusive Rayleigh number $\Ra_D$. The two numbers are related by a simple formula, viz., $\Ra_D = \Ra_\gamma^{3/2} (1-\gamma/f_0)^{-5} $.

We show that when $\Ra_\gamma$ is below a critical value, conductive and viscous dissipation of heat and momentum respectively is sufficient to have a motionless steady-state solution for the fluid with all the heat transfer occurring through thermal conductivity alone. Steady-state linear stability analysis for small perturbations to this steady-state was conducted to identify the critical $\Ra_\gamma$ for transition from conductive state to a convective state. It was found that for small values of $\gamma/f_0$, the least stable horizontal mode has a long wavelength (wavenumber $\kappa \to 0$) and the value can be appproximated by a simple formula given in eqn. \eqref{eq:smallkRac}. For larger $\gamma/f_0$, the critical $\Ra_\gamma$ is given by a constant value of $78$. The critical $\Ra_\gamma$ from 2D simulations showed an excellent match with the theoretically obtained values by solving the linear eingenvalue problem resulting from the stability analysis. 

Beyond this critical value, the fluid no longer remains motionless and is unstable to small perturbations which induce a convective motion in the system. 
We have further shown that in this convective regime, the scaling of important measured quantities such as the Nusselt number, the non-dimensionalised mass-flux and average velocity in updrafts with $\Ra_\gamma$ converge to a single power-law, with some dependence of the magnitude on the dimensionless ratio $\gamma/f_0$. For large enough $\Ra_\gamma > \sim 10^4$, the heat-flux and the mass-flux scale as $\Ra_\gamma^{0.5}$ and $\Ra_\gamma^{0.7}$ respectively. 

The bulk cooling term also introduces an up-down asymmetry in the fluid - this asymmetry is particularly marked when the flow is only weakly convective, ie. $\Ra_\gamma$ is only slightly larger than the critical value. For strongly convective flows, the degree of the asymmetry as measured by the volume of the domain occupied by updrafts (regions where $w>0$) and the skewness of the vertical velocity decreases, where nearly half the domain is made up of updrafts and the skewness goes close to $0$. While the 3D flows are identical to the 2D flows in the scaling of heat and mass transfers with $\Ra_\gamma$, they retain a large degree of up-down asymmetry even at $\Ra_\gamma \sim 10^5$. 

In this study, we have added to the existing body of literature on various models of thermal convective systems which help us understand atmospheric convection under different conditions. While the system has been introduced elsewhere, this is to the best of our knowledge the first study that systematically varies the parameters for a fixed-flux, internally cooled, stratified convective system, a system with particular relevance to the sub-cloud atmospheric boundary layer. Our main aim here has been to demonstrate the simple scaling of the system with respect to various input parameters and provoke future studies which explore higher Rayleigh numbers as well as mixed boundary conditions and non-uniform radiative cooling. 

Idealised models of convection, including those with moisture and water phase changes serve to provide a tool to study the complex atmospheric system in a simplified setting where it is far easier to delineate the dynamic effects and feedbacks due to individual processes. In the future, we plan to study idealised models which include moist dynamics as well as global-scale idealised models which include the effects of rotation. 

\section*{Acknowledgments}
This project has received funding from the European Union's Horizon 2020 research and innovation programme under the Marie Sklodowska-Curie grant agreement No. 101034413.  CM gratefully acknowledges funding from the European Research Council (ERC) under the European Union’s Horizon 2020 research and innovation program (Project CLUSTER, Grant Agreement No. 805041).

\section*{Conflict of Interests and Data Availability}
The authors declare no conflicts of interest. Datasets and the code used to generate the data are available from the authors upon reasonable request. 

\appendix
\section{The heat-equation}\label{app:heateq}
We start with the first law of thermodynamics in enthalpy form given by 
\begin{equation}
    c_p \frac{D T}{Dt} - \frac{1}{\rho} \frac{D p}{D t} = \Dot{Q}
\end{equation}
where %$m$ is the mass of the fluid parcel, %CM: remove (there is no m)
$c_p$ is the specific heat-capacity at constant pressure and $\Dot{Q}$ is the rate of diabatic heating or cooling per unit mass. 
We make the assumption ($D_t p \approx w \partial_z p_{ref} \approx - \rho_{ref} g w $), which is the condition of hydrostatic balance. This also assumes that a given parcel always has the same pressure as its environment, which is a common assumption made in parcel theory. This gives,

%Using the condition of hydrostatic balance ($D_t p \approx - \rho g w $) gives the heat-equation 

\begin{equation}
    \partial_t T + \uu \cdot \nab T + (g/c_p) w = \Dot{Q}/c_p,
\end{equation}

where the RHS includes thermal dissipation as well as diabatic cooling, which is given by $-R$ in this study. $g/c_p \equiv \gamma$  is the well known dry adiabatic lapse-rate.

\section{Thermal and Viscous dissipation}
\subsection{Thermal Dissipation}\label{AppendixA}
The thermal dissipation is defined as 
\begin{equation}
    \epsilon_T \equiv \kappa \bigl\langle(\partial_i T(\boldsymbol{x},t))^2\bigr\rangle_V.
\end{equation}

Following \citep{siggia1994high}'s analysis for Rayleigh-B\'enard convection, we multiply equation \eqref{eq:Heat-eqn} by temperature $T$ and average the product over the entire domain and time to give
\begin{multline}
    \frac{1}{2}\frac{d \langle T^2 \rangle}{dt} + \frac{1}{2} \bigl\langle \uu \cdot \nab (T^2) \bigr\rangle + \gamma  \bigl\langle w T \bigr\rangle + \bigl\langle R T \bigr\rangle   \\
    = \kappa \bigl\langle T \nabla^2 T \bigr\rangle = \kappa \bigl\langle \nab \cdot (T \nab T) \bigr\rangle - \kappa \bigl\langle | \nab T |^2 \bigr\rangle,
    \label{eq:Tdot_heat}
\end{multline}
Since we working in the stationary regime, ($\partial_t \langle \cdot \rangle = 0$). Further, using the incompressibility condition (eq. \eqref{eq:incomp}) and the fluid rigid boundary condition (\eqref{eq:fluigbound}), we get
\begin{equation}
    \bigl\langle \uu \cdot \nab (T^2) \bigr\rangle_V 
    = \bigl\langle \nab \cdot (\uu T^2)\bigr\rangle_V = 0.
\end{equation}
Then, equation \eqref{eq:Tdot_heat} becomes 
\begin{equation}
    \kappa \bigl\langle | \nab T |^2 \bigr\rangle = \kappa \bigl\langle \nab \cdot (T \nab T) \bigr\rangle 
    - \gamma \bigl\langle w T \bigr\rangle - R \bigl\langle  T \bigr\rangle,
\end{equation}
or 
\begin{equation}
    \epsilon_T = \kappa \bigl\langle \nab \cdot (T \nab T) \bigr\rangle 
    - \gamma \bigl\langle w T \bigr\rangle - R \bigl\langle  T \bigr\rangle.
\end{equation}
Using the Gauss theorem, the first term of $\epsilon_T$ can be written in terms of a surface integral as
\begin{equation}
    \kappa \bigl\langle\nab \cdot (T \nab T) \bigr\rangle 
    = \frac{\kappa}{L_z} \biggl[ \Bigl\langle T \partial_z T \Bigr\rangle_{z=L_z} 
    - \Bigl\langle T \partial_z T \Bigr\rangle_{z=0}\biggr].
\end{equation}
Setting $\partial_z T|_{z=0} = -f_0$ and $\partial_z T|_{z=L_z} = -f_1 = 0$ gives the expression in eqn. \eqref{eq:heat_disip}
\begin{equation}
    \kappa \langle |\nab T|^2 \rangle =  \frac{\kappa}{L_z} (T_0 f_0) - \gamma \langle wT \rangle - R \langle T \rangle. 
\end{equation}

\subsection{Viscous Dissipation}\label{AppendixB}
Taking the dot product of eq.~\eqref{eq:Nav-Stokes} with $\uu$ and taking the statistical average over the whole volume as above, we get
\begin{multline}
    \frac{1}{2} \frac{d}{dt} \bigl\langle (\uu \cdot \uu ) \bigr\rangle 
    + \frac{1}{2} \bigl\langle \uu \cdot \nab (\uu \cdot \uu ) \bigr\rangle \\
    = - \bigl\langle \uu \cdot \nab p \bigr\rangle + \nu \bigl\langle \uu \cdot \nabla^2 \uu \bigr\rangle + \beta g \bigl\langle w T \bigr\rangle.
    \label{eq:udot_fluid}
\end{multline}
In the stationary state, the terms of the form $d_t \langle \cdot \rangle$ vanish. Using the incompressibility condition and the rigid boundary condition, we have
\begin{equation}
    \bigl\langle \uu \cdot \nab (\uu \cdot \uu ) \bigr\rangle_V = \Bigl\langle \nab \cdot \bigl[\uu (\uu\cdot\uu)\bigr] \Bigr\rangle_V = 0
\end{equation}
\begin{equation}
    \langle \uu \cdot \nab p \rangle_V = \bigl\langle \nab \cdot (\uu p) \bigr\rangle_V = 0
\end{equation}
\begin{equation}
\begin{split}
    \langle \uu \cdot \nabla^2 \uu \rangle_V 
    &= \frac{1}{2} \bigl\langle \nab \cdot \nab (\uu \cdot \uu) \bigr\rangle_V - \sum_{i,j} \biggl\langle \Bigl ( \frac{\partial u_j}{\partial x_i} \Bigr) ^2 \biggr\rangle_V\\
    &= - \sum_{i,j} \biggl\langle \Bigl ( \frac{\partial u_j}{\partial x_i} \Bigr ) ^2 \biggr \rangle_V\\
    &= -\frac{1}{2} \sum_{i,j} \biggl\langle \Bigl( \frac{\partial u_i}{\partial x_j} + \frac{\partial u_j}{\partial x_i} \Bigr) ^2 \biggr\rangle_V.
\end{split}
\end{equation}
So, equation \eqref{eq:udot_fluid} becomes 
\begin{equation}
    \frac{\nu}{2} \sum_{i,j} \biggl \langle \Bigl( \frac{\partial u_i}{\partial x_j} + \frac{\partial u_j}{\partial x_i} \Bigr) ^2 \biggr\rangle = \beta g \langle w T \rangle,
\end{equation}
or 
\begin{equation}
    \epsilon \equiv \frac{\nu}{2} \sum_{i,j} \biggl \langle \Bigl( \frac{\partial u_i}{\partial x_j} + \frac{\partial u_j}{\partial x_i} \Bigr) ^2 \biggr\rangle= \beta g \langle w T \rangle.
\end{equation}


\bibliographystyle{elsarticle-num-names}
\bibliography{refs}
\end{document}